# A Fast Deep Learning Approach for Beam Orientation Optimization for Prostate Cancer Treated with Intensity Modulated Radiation Therapy

Azar Sadeghnejad Barkousaraie, Olalekan Ogunmolu, Steve Jiang, and Dan Nguyen[*]

Medical Artificial Intelligence and Automation (MAIA) Laboratory, Department of Radiation Oncology, UT Southwestern Medical Center, Dallas, TX

## Abstract

**Purpose:** Beam orientation selection, whether manual or protocol-based, is the current clinical standard in radiation therapy treatment planning, but it is tedious and can yield suboptimal results. Many algorithms have been designed to optimize beam orientation selection because of its impact on treatment plan quality, but these algorithms suffer from slow calculation of the dose influence matrices of all candidate beams. We propose a fast beam orientation selection method, based on deep learning neural networks (DNN), capable of developing a plan comparable to those developed by the state-of-the-art column generation method. Our model's novelty lies in its supervised learning structure (using column generation to teach the network), DNN architecture, and ability to learn from anatomical features to predict dosimetrically suitable beam orientations without using dosimetric information from the candidate beams. This may save hours of computation.

**Methods:** A supervised DNN is trained to mimic the column generation algorithm, which iteratively chooses beam orientations one-by-one by calculating beam fitness values based on Karush-Kush-Tucker optimality conditions at each iteration. The DNN learns to predict these values. The dataset contains 70 prostate cancer patients—50 training, 7 validation, and 13 test patients—to develop and test the model. Each patient's data contains 6 contours: PTV, body, bladder, rectum, and left and right femoral heads. Column generation was implemented with a GPU-based Chambolle-Pock algorithm, a first-order primal-dual proximal-class algorithm, to create 6270 plans. The DNN trained over 400 epochs, each with 2500 steps and a batch size of 1, using the Adam optimizer at a learning rate of $1 \times 10^{-5}$ and a 6-fold cross-validation technique.

**Results:** The average and standard deviation of training, validation, and testing loss functions among the 6-folds were 0.62±0.09%, 1.04±0.06%, and 1.44±0.11%, respectively. Using column generation and supervised DNN, we generated two sets of plans for each scenario in the test set. The proposed method took at most 1.5 seconds to select a set of five beam orientations and 300 second to calculate the dose influence matrices for 5 beams and finally 20 seconds to solve the fluence map optimization. However, column generation needed around 15 hours to calculate the dose influence matrices of all beams and at least 400 seconds to solve both the beam orientation selection and fluence map optimization problems. The differences in the dose coverage of PTV between plans generated by column generation and by DNN were 0.2%. The average dose differences received by organs at risk were between 1 and 6 percent: Bladder had the smallest average difference in dose received (0.956±1.184%), then Rectum (2.44±2.11%), Left Femoral Head (6.03±5.86%), and Right Femoral Head (5.885±5.515%). The dose received by Body had an average difference of 0.10± 0.1% between the generated treatment plans.

**Conclusions:** We developed a fast beam orientation selection method based on a DNN that selects beam orientations in seconds and is therefore suitable for clinical routines. In the training phase of the proposed method, the model learns the suitable beam orientations based on patients' anatomical features and omits time intensive calculations of dose influence matrices for all possible candidate beams. Solving the fluence map optimization to get the final treatment plan requires calculating dose influence matrices only for the selected beams.

[*] Corresponding author: Dan.Nguyen@UTSouthwestern.edu

# I.    Introduction

External beam radiation therapy (EBRT) is a common treatment method for various types of cancers. EBRT uses a machine to emit high-energy radiation to the patient's body to damage cancerous cells; but these radiation beams, while traveling through the body, do not distinguish between healthy and cancerous cells, so they ultimately damage healthy tissue and critical structures as well. Too much damage to healthy critical structures degrades patients' quality of life and should be minimized as much as possible. However, the integral dose to the patient's body tends to be fairly constant regardless of the planning modality and expertise,[1-5] meaning that the best a planner can do is to decide where the excess radiation should be positioned. Finding a treatment plan that maximizes the delivery of the prescription dose to cancerous cells while minimizing toxicity and undesirable endpoints to healthy organs is a challenging problem and is the main purpose of the treatment planning process.

Intensity Modulated Radiation Therapy (IMRT)[6-9] is a common method of EBRT that delivers radiation beams with various intensities and from different static directions towards the planning target volume (PTV). The configuration of beam directions has a major effect on the quality of the treatment plan and can be considered as a large scale combinatorial optimization problem,[10] usually called a beam orientation optimization (BOO) problem. BOO has been studied since 1967[11]; however, in the current treatment planning workflow, the beam direction is selected manually by the planner, following a time-consuming trial-and-error process that typically yields suboptimal solutions.[12,13] To accurately measure the impact of the BOO solution in the dosimetric space, a fluence map optimization (FMO)[14] problem must be solved. FMO is the problem of finding the optimal intensity of beam profiles to generate a high quality plan.[12] To calculate beam profiles, each beam is considered as a collection of small beamlets whose intensities can be controlled individually.[15] The number of candidate beams can be very large; for example, there are 180 candidate beams for a coplanar geometry with 2-degree separation and 1162 candidate beams for a noncoplanar geometry with 6-degree separation.

Modern BOO methods typically solve the problem in the radiation dose domain, which requires precomputing the dose influence matrices for all candidate beam orientations and then solving the FMO. The final objective function in these works is usually a function of the differences between the actual and prescribed dosage received by heathy tissues, organs at risk (OARs), and the PTV.[16-28] But computing the dose influence matrices and the FMO are both very complex and time intensive operations,[29] taking hours to compute dose influence matrices and minutes to solve the FMO, which ultimately hampers the implementation of BOO in clinical routines.[13,20,24,30-32] Breedveld et al.[13] used a "wish list" to prioritize constraints and objective functions related to OARs and the tumor and to iteratively add new beam orientations to the set of currently selected beams. Yarmand et al.[31] scored beamlets of candidate beams based on their contributions to irradiating tumors and weighted the average of the total dose received by OARs. Then, for each candidate beam, they generated L+2 (L number of OARs) apertures with the maximum sum of the beamlets' contribution scores. Next, they defined and solved an ideal plan that minimized the dosages delivered to OARs. Finally, they used another model to minimize the number of beams in the plan while using the objective function of the ideal plan as a constraint with $\varepsilon$ distance to the optimal solution. Later, Yarmand et al.[33] elaborated on their previous work[31] and offered three heuristic approaches to reduce the computation time of the proposed model: neighbor cuts, beam elimination, and beam reduction. Cabrera et al.[12] proposed a two-phase strategy in which they found non-dominated solutions for locally optimal beam orientations by first solving a deterministic local search and then generating multiple FMO solutions for each set of beams. The final Pareto optimal solution consisted of beam sets that had at least one non-dominated FMO solution. Craft et al.[34] proposed an interactive method to navigate multiple Pareto surfaces with a Piercing scalarization method, which allows users to choose the objective function that they want to improve, and then the method searches the set of beam orientations and their Pareto surfaces to find the best or closest solution. Cabrera et al.[35] presented Pareto local search and adaptive Pareto local search approaches as two multi-objective local search problems. They used a biological model objective function to help spare the OARs.

To overcome the problem of time intensive calculations for dose-based metrics, several researchers have used purely anatomical metrics for BOO. Cho et al.[36] introduced the concept of target eye view maps for conformal radiotherapy treatment planning. Meyer et al.[37] sorted potential beam orientations based on the distance to the PTV and OARs in increasing order, then selected beams with the same order, subject to a minimum distance threshold, for 3D-CRT planning for prostate, brain, and sinus cancer. Potrebko et al.[29] addressed IMRT and 3D-CRT BOO problems for both coplanar and noncoplanar cases. First, they located beam orientations that bisected the target and OARs to spare critical structures; then, they located beams parallel to 3D flat surface features of the PTV to facilitate conformity of target coverage. Llacer et al.[38] used a combination of computer vision, beam's eye view techniques, and a neural network to define a new cranial lesion treatment planning method for coplanar and noncoplanar beams, based exclusively on geometric information. Amit et al.[39] used clinically approved treatment plans to train a random forest regression algorithm to learn the relation between the patient's anatomy and beam orientations. They proposed sets of beam-dependent and beam-independent anatomical features (such as overlapping between the tumor and each OAR in beam's eye view, and location of the tumor, respectively) to solve coplanar beam IMRT for thoracic cancer. Although using pure anatomical metrics would improve the computation time, there is no guarantee about the quality of the solution, because none of these methods considers the FMO solution, which is the most accurate metric for BOO solutions.

In contrast, artificial intelligence (AI) is an attractive tool for solving the BOO problem, given its superior speed and promising results for medical applications. The vast employment of machine learning algorithms to solve dose optimization,[40-42] image segmentation,[43-45] outcome prediction,[41,46] and quality assurance problems in recent years indicate the success of AI applications in radiation oncology and medical physics.[47] Specifically, the success of algorithms such as convolutional neural networks (CNN)[48] in image processing and the learning capability of modern machine learning techniques enable treatment planners to provide patient-specific plans based on the patient's anatomical features and by learning from complicated optimization methods or physician behaviors.

This work aims to develop a fast and flexible BOO method based on deep neural networks (DNN) that provides a solution in a matter of seconds and that, therefore, can be directly implemented in clinical routine to accelerate the treatment planning process for patients with cancer. The proposed DNN learns the connection between the patient's anatomy and the optimal set of beam orientations from the patient's anatomical features and an optimization algorithm, and it has the attractive feature of predicting a set of beam orientations without prior knowledge of dose influence matrix values. Although this study uses only one specific objective function and optimization algorithm, the general structure of the method is applicable to any objective function defined for BOO and any iterative optimization algorithm.

The images of 70 clinical prostate cancer patients were augmented by rotation and a random OAR weight generator to form 30800 input samples, then used to train, evaluate, and test the proposed DNN. The optimization algorithm used to find the optimal set of beams to train the model is a greedy iterative algorithm known as column generation (CG). CG algorithms such as the one in this study have demonstrated success in finding the best treatment plan.[22,49-55] The proposed CG method starts with an empty set of beams, and in each iteration, CG selects one beam with the highest potential to improve the current solution and adds it to the set, and the respective FMO is solved. We utilized the Chambolle-Pock algorithm,[56-58] a first-order primal-dual proximal-class algorithm, on the GPU to solve the FMO problem. We then train the DNN to learn the beam orientation reasoning of CG. By learning to mimic CG, which has considered the FMO solution, the DNN will also internalize this FMO solution, even though it is not directly using it. To test the feasibility of the framework, we considered a total number of five beams to treat the patient.

Specifically, in this study, the plan optimization approach consists of an iterative scheme where the column generation phase selects the most favorable beam to add to the problem, followed by an FMO optimization phase for the dose distribution. Based on this result, the CG phase will then be able to suggest a new beam orientation for inclusion. In the DNN approach, the CG phase is replaced by DNN, which does not require the computation of the beam influence matrices and used the same iterative structure to select predefined number of beam orientations. The proposed approach is detailed in section II, the computational results of using DNN compared to CG are presented in section **Error! Reference source not found.**, while discussion and conclusions are included in sections **Error! Reference source not found.** and **Error! Reference source not found.**, respectively.

## II.  Materials and Methods

AI's ability to solve complicated problems in a short time makes it an attractive tool for solving time-sensitive problems, such as BOO. The proposed method benefits from AI's capability to learn complicated behaviors and CNN's abilities in image processing, and it predicts a suitable and high quality beam orientation configuration given a certain state of the problem. The state of the problem is defined by patient images, with respect to the patient contours and organ importance weights, and the currently selected set of beam orientations (B). The prediction effectively determines the next best beam to add to the solution set. This method has an iterative structure in which a deep learning neural network (DNN) is used repeatedly to predict the next beam to be added to B. For example, to select 5 beam orientations, the proposed DNN will be called five times consecutively, starting with $|B| = 0$ and stopping when $|B| = 5$. Details of this iterative approach are presented in this section.

Here, we propose a supervised deep learning algorithm with convolutional layers to solve BOO problems. BOO can benefit from the proposed method not only because of its fast prediction time, but also because it can detect the anatomical features of the patient's body. The proposed network starts with the patient's anatomy, including the structure segments and organ importance weights, and a set of already chosen beam orientations, and it predicts the next beam orientation to be selected. The detailed explanation of the network structure is provided in section II.A. This network predicts the next beam to be added to the solution set and should be iteratively called to reach the user-defined number of beam orientations. The problem that is solved in each iteration, called a Limited BOO problem, is presented in section II.C. The solutions to a set of known problems are required to train a supervised learning network. The method used in this paper and its iterative structure are presented in section II.C.1.

### II.A    Network Structure

**Figure 1** presents the structure of the proposed network. The model's overall structure is an encoder-decoder–style network. This network has two inputs: an anatomical input (*Input 1*) and a beam orientation input (*Input 2*). The anatomical input consists of two channels, each of size $64 \times 64 \times 64$. The first channel represents PTV as a target channel, and the second channel represents weighted OARs as an avoidance channel. The channels have data similar to binary masks of the structures, but instead of ones, The PTV and structures are assigned non-zeros values. This non-zero value is the structure weighting that is assigned during the optimization step, which is described in detail in **Section II.D**.

The second input is a Boolean array of size 180 (the number of candidate beam orientations at 2-degree resolution), which represents beam orientations that are already chosen as ones and the rest as zeros. The second input and its complementary[a] array (as $B$ and $B^C$ respectively) are used throughout the network to b represent the set of already chosen orientations. In our experience, using two complementary Boolean arrays, instead of just one, enhances the influence of already selected beam orientations.

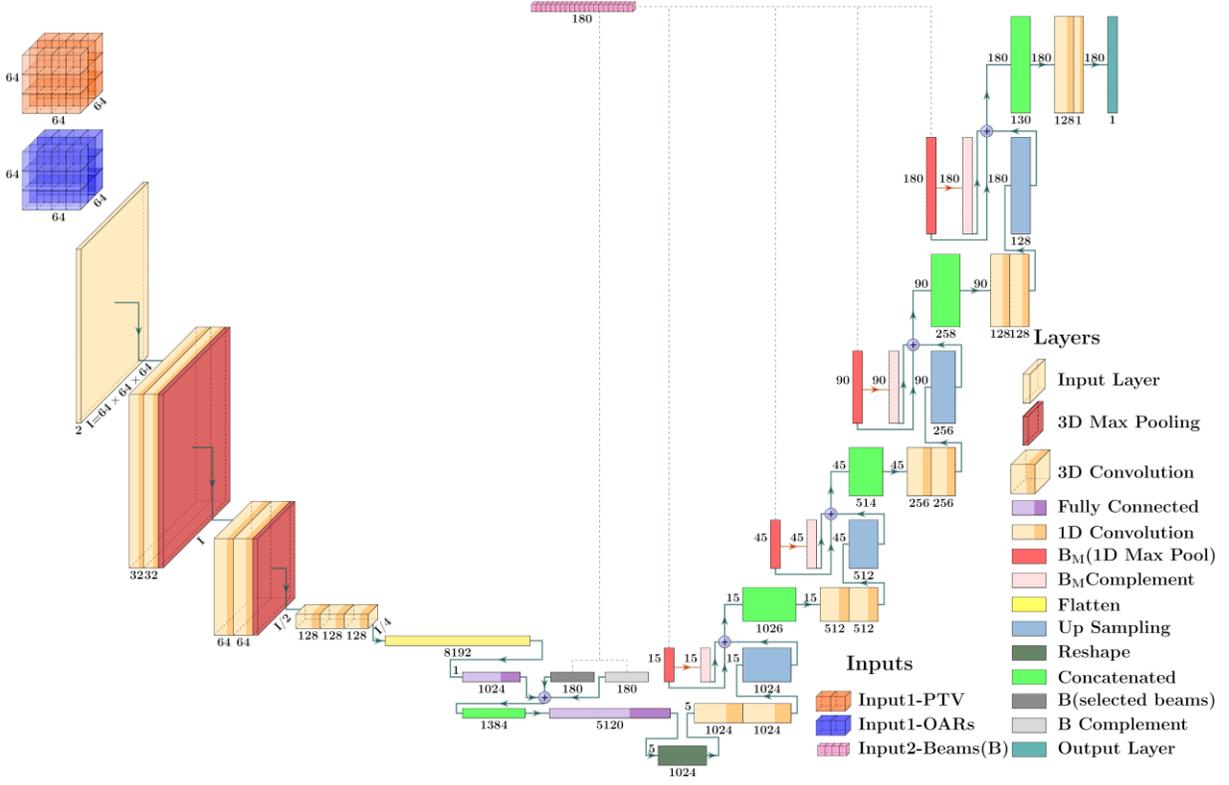

**Figure 1.** *Proposed Deep Learning Neural Network Structure.*

The network has three blocks of layers. The first block analyzes the patient's anatomy using 3-dimensional convolutional layers. The first two levels of the block contain two consecutive 3-dimensional convolutional layers with $5 \times 5 \times 5$ kernel sizes and one max pooling layer of size $2 \times 2 \times 2$, represented by red tensors in **Figure 1**. The third level uses three consecutive 3-dimensional convolutional layers with $3 \times 3 \times 3$ kernel sizes; the output of the third convolutional layer will be flattened to a 1-dimensional layer (yellow layer in the picture), and, after a fully connected layer (purple layer), the size will be reduced to 1024 features. The second block of layers adds the second input (beam orientation array) as two arrays $B$ and $B^C$, shown as dark and light gray layers, respectively, to the 1-dimensional fully connected anatomy layer. Then the concatenated layer (light green layer) will be used as input to another fully connected layer with output size of 5120 features, which will be reshaped to 5 rows of 1024 features (shown in dark green color). The third block of layers has five stages, each of which starts with two 1-dimensional convolutional layers with kernel size 3, then an up sampling layer, max pooling of the beam information vector as $B_M$ (red 1-dimensional layer) and its complement $B_M^C$ (pink 1-dimensional layer), and the concatenation layer of up sampled, max pooled beam, and its complement. After the last convolution layer of the last stage, there is only one feature array of size 180, which represents the dual values of candidate beam orientations. Regarding the max pooling of the beam information vector, the pool size dynamically changes based on the size of the current stage's data. Specifically, the pooling size is 12, 4, 2 and 1, respectively. The dual value of beam $b$ shows the potential improvement in the current state of the problem (the value of the objective function), if beam $b$ is added to the current set of selected beams ($B$). We used the inversed and normalized dual array as the probability distribution of beam orientation selection, given the current state of the problem.

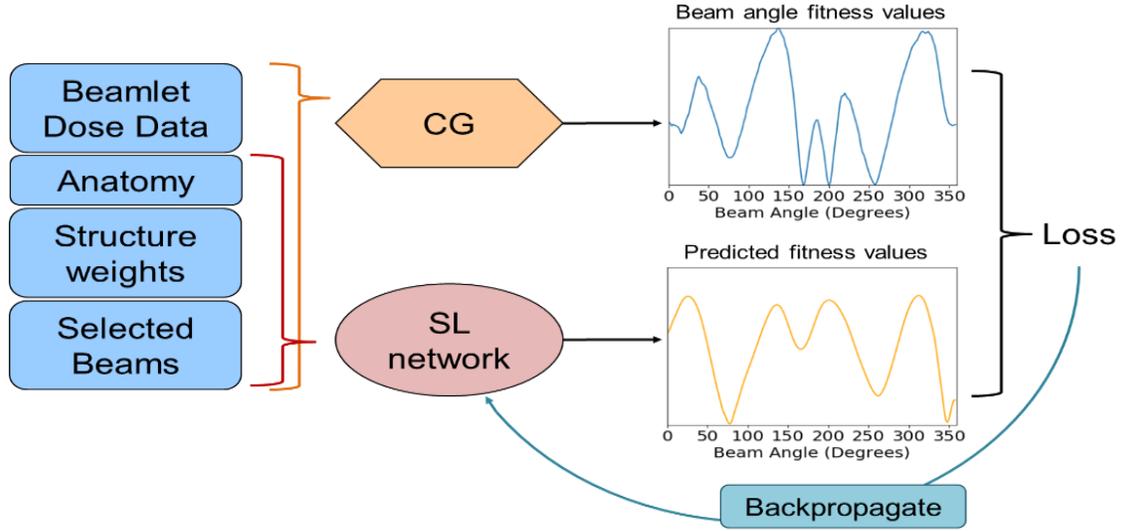

**Figure 2.** *Schematic of the Supervised Training Structure to Predict Beam Orientation Fitness Values. Column Generation (CG) as teacher and Supervised Learning (SL) Network as Trainee.*

The activation functions of all the layers in the proposed model are Scaled Exponential Linear Unit (SELU). SELU is a self-normalizing activation function that converges to zero mean and unit variance even in the presence of noise, and it makes learning robust even for DNNs with many layers.[59] The loss function of the model is Mean Squared Error (MSE) with Adam optimizer[60] with a learning rate of $1 \times 10^{-5}$.

## II.B Supervised Structure

The goal of a Supervised Learning Neural Network (SLNN) is to find network weights such that the SLNN output has the minimum difference from teacher-given, real-valued labels or targets. In the proposed method, the teacher is the CG algorithm, and the target labels are fitness values, $f$, calculated from the KKT conditions, defined in equation (9). A schematic of the proposed supervised training structure is depicted in **Figure 2**. The sections that follow describe the optimization problem and the column generation algorithm to solve this problem.

## II.C Full and Limited BOO Problem

From an optimization perspective, the fluence map optimization problem, $P(B)$, can be written as

$$\underset{x,y}{minimize} \qquad F(y) \qquad\qquad (1)$$

$$subject\ to \qquad y = \sum_{b \in B} \begin{bmatrix} D_{b,s=s_1} \\ \vdots \\ D_{b,s=s_T} \end{bmatrix} x_b \qquad\qquad (2)$$

$$x_b \geq 0 \qquad\qquad for\ b \in B \qquad\qquad (3)$$

where $F$ is the objective function and $D_{b,s}$ is the dose influence matrix for the $b^{th}$ beam and the $s^{th}$ structure. $B$ is the set of selected beam orientations. Assuming no time, computation, or delivery constraints, the single best treatment plan would use all possible candidate beams to treat the patient. In this case, we would define the master problem by setting $B = B_{all}$ and solving $P(B_{all})$, where $B_{all} = \{b_j : \forall j \in J\}$ is the set of all available beam orientations in $B_{all}$. $S$ is the set of available structures (PTV, OARs, tuning structures). While this formulation would theoretically give the best plan, it is not feasible to deliver this plan to the patient[c]. Instead, we look at the limited problem, $P_{limit}(B_{all}, n)$, written as:

$$\underset{x, y, B_{limit}}{minimize} \qquad F(y) \qquad\qquad (4)$$

$$subject\ to \qquad y = \sum_{b \in B_{limit}} \begin{bmatrix} D_{b,s=s_1} \\ \vdots \\ D_{b,s=s_T} \end{bmatrix} x_b \qquad\qquad (5)$$

$$x_b \geq 0 \qquad\qquad for\ b \in B_{all} \qquad (6)$$

$$|B_{limit}| \leq n \qquad\qquad B_{limit} \in B_{all} \qquad (7)$$

where $B_{limit}$ is a set of limited beam orientations, and the number of beams in $B_{limit}$ must be restricted to less than a user-defined number of beams, $n$. For a reasonable $n$ (e.g. $n=5$ or 10), the limited formulation is feasible for treatment delivery to the patient; however, as mentioned in the introduction, solving for the optimal set of beams, $B_{limit}^*$, is challenging, and an exhaustive search is infeasible for a large set of candidate beams, $|B_{all}|$.

We have chosen a greedy algorithm based on column generation (CG) to solve the problem. CG has been shown to be effective for beam orientation optimization in radiation therapy problems.[61-63] While CG does not guarantee an optimal solution, the CG algorithm has been shown to have superior performance when compared to clinical plans [10,30,50,53,61,63-73], and the algorithm is summarized in the sections that follow. A list of all indices, parameters, variables, and functions used in this paper and their definitions are provided in **Table 1**.

### II.C.1  Column Generation

CG approximates the solution to the limited BOO problem by iteratively adding beams one at a time to $B_{limit}$ until $|B_{limit}| = n$ or until optimality to the master problem has been reached. During each iteration, the next beam to be added is the beam that best minimizes the immediate objective value. In optimization terms, CG attempts to solve the limited problem, $P_{limit}(B_{all}, n)$, by solving the more general problem, $P(B)$, through the iteration. This concept is presented in **Algorithm 1**.

---

**Algorithm 1.** *Brief Column Generation Structure for Beam Orientation Selection*

---

1. Initialize $B_{limit}$ as an empty set: $k = 0$, $B_{limit}^0 = \emptyset$

2. While $|B_{limit}| < n$:

   a. $b^{k+1} = \underset{\tilde{b} \in B_{all} \backslash B_{limit}^k}{argmin} \{ P(\hat{B}) : \hat{B} = B_{limit}^k + \tilde{b},\ |\hat{B}| = |B_{limit}^k| + 1 \}$

   b. $B_{limit}^{k+1} = B_{limit}^k + b^{k+1}$

---

where $k$ is the iteration number. The problem in its current form would still be expensive to evaluate since it would require a fluence map optimization for each beam in $B_{all} \backslash B_{limit}^k$ at each iteration. However, this can be bypassed by evaluating the Karush-Kuhn-Tucker (KKT) conditions for optimality[74,75] of the master problem, $P(B)$, during each iteration. If all KKT conditions are met, then the problem is considered optimal. If not, it is possible to check which variables are furthest from optimality and would, therefore, improve the objective value the quickest if corrected. For our case, minimizing $P(B_{limit})$ but evaluating KKT conditions for $P(B_{all})$ would reveal that most, if not all, of the beams in the set $B_{all} \backslash B_{limit}$ are not optimal (i.e. if selected, would improve the solution). And the KKT conditions reveal which single beam would best improve the objective value at the next iteration.

**Table 1.** *List of Indices, Parameters, Variables and Functions*

| Indices and Parameters | |
|---|---|
| $t \in \{1, 2, \dots, T\}$ | Structure index |
| $S = \{s_t : \forall t\}$ | Set of available structures (PTV, OARs, and tuning structures) |
| $s_t$ | Set of voxels in $t^{th}$ structure |
| $w_s$ | User-defined weight for structure $s$ |
| $p_s$ | Prescription dose for structure $s$ |
| $B$ | Set of selected beam orientations to solve FMO |
| $P(B)$ | Fluence map optimization problem of set $B$ |
| $B_{all} = \{b_j : \forall j \in J\}$ | Set of all available beam orientations |
| $J$ | Set of all indices in $B_{all}$ |
| $b_j$ | Set of beamlets in $j^{th}$ beam |
| $D_{b,s}$ | Dose influence matrix for beam $b \in B$ and structure $s \in S$ |
| $n$ | Maximum number of beam orientations in $B_{limit}$ |
| $n_B$ | Number of beam orientations in set $B$, $n_B = |B|$ |
| $P_{limit}(B_{all}, n)$ | FMO problem of $n_B$ number of beam orientations in set $B_{all}$, where $n_B \leq n$ |
| $B_{limit} \subseteq B_{all}$ | A set of limited beam orientations |
| $k$ | Iteration number in column generation algorithm |
| $B_{limit}^k$ | Set of selected beam orientations in iteration k of CG algorithm |
| $b_k$ | The one beam selected in iteration k of CG algorithm |
| **Variables** | |
| $x_b$ | The intensity array of the candidate beam $b \in B$ |
| $x_{b,i}$ | Element of array $x_b$, intensity of $i^{th}$ beamlet of beam $b \in B$ |
| $y = \begin{bmatrix} y_{s=s_1} \\ \vdots \\ y_{s=s_T} \end{bmatrix}$ | An arrays of size $T$, with elements $y_{s \in S}$, where $y_s$ is the value array of the total dose received by voxels in structure $s$ |
| $z$ | Lagrange multipliers of size $T$ |
| $z_s$ | Array of Lagrange multiplier of voxels in the structure $s$ |
| $v$ | Lagrange multipliers of size $n_B$ |
| $v_b$ | Lagrange multiplier of beamlets in candidate beam orientation $b$ |
| $v_{b,i}$ | Lagrange multiplier for $i^{th}$ beamlet of candidate beam orientation $b$ |
| $r_b$ | The possible improvement in objective function (reduced cost) by adding $b$ to $B$ |
| **Functions** | |
| $F$ | Objective function |
| $L(x, y, z, v)$ | Langrangian function of $P(B_{all})$ problem |

To obtain the KKT conditions, we first write out the Lagrangian, $L(x, y, z, v)$, for the master problem, $P(B_{all})$:

$$L(x, y, z, v) = F(y) + \langle z, \sum_{b \in B} \left( \begin{bmatrix} D_{b,s=s_1} \\ \vdots \\ D_{b,s=s_T} \end{bmatrix} x_b \right) - y \rangle - \sum_{b \in B_{all}} \langle v_b, x_b \rangle \tag{8}$$

where $z$ and $v \geq 0$ are the Lagrange multipliers associated with their respective constraints (5) and (6). The KKT conditions are derived from this Langrangian and can be divided into four categories, shown in **Table 2**.

In particular, we are interested in the dual feasibility condition, constraint (14) in **Table 2**. We expect that this condition will not be met for $b \in B_{all} \backslash B_{limit}^k$ when solving the subproblem for $b \in B_{limit}^k$ for the $k^{th}$ iteration. To calculate this, we first obtain $z \in \partial F(y)$, which is automatically acquired when solving the problem $P(B_{limit}^k)$ using a primal-dual algorithm. Then, all $v_b$ can be calculated using the stationarity condition, as represented in constraint (10). The beam where the value $\sum_i (v_{b,i})_-$ is the most negative (a.k.a. furthest from meeting the optimality criterion) would improve the objective value the most at the next iteration if added to $B_{limit}$. Letting $r_b = -\sum_i (v_{b,i})_-$, we then define our fitness vector, $f$, as:

$$f = \frac{\begin{bmatrix} r_{b=b_1} \\ \vdots \\ r_{b=b_{|J|}} \end{bmatrix} - \min(r)}{max(r) - \min(r)} \tag{9}$$

Here, the fitness is normalized from zero to one, which is set up for training the neural network later on. The larger the value of the fitness, the better the beam is for improving the objective value at the next iteration. We can find the next best beam $b^{k+1} = argmax(f^k)$. **Algorithm 2.** describes how the CG algorithm uses the KKT conditions to find a suitable set of beams.

**Table 2.** *KKT Conditions for problem* $P(B_{all})$

| **KKT Conditions:** | | |
|---|---|---|
| *Stationarity* | $v_b = \begin{bmatrix} D_{b,s=s_1} \\ \vdots \\ D_{b,s=s_T} \end{bmatrix}^T z$ | $for\ b \in B_{all}$ (10) |
| | $z \in \partial F(y)$ | (11) |
| *Primal feasibility* | $y = \sum_{b \in B} \begin{bmatrix} D_{b,s=s_1} \\ \vdots \\ D_{b,s=s_T} \end{bmatrix} x_b$ | (12) |
| | $x_b \geq 0$ | $for\ b \in B_{all}$ (13) |
| *Dual feasibility* | $v_b \geq 0$ | $for\ b \in B_{all}$ (14) |
| *Complementary slackness* | $v_{b,i} x_{b,i} = 0$ | $\forall b, i$ (15) |

---

**Algorithm 2.** *Solving a Sequence of a Limited BOO Problem to Select n Beam Orientations*

---

1. Create a one-dimensional array (A) with the same size as $J$ (in this work $|J| = 180$)
2. Set current number of selected beam orientation (B) as 0, $n_B = 0$
3. Initialize array A with zeros
4. Set the value of current objective function $F_{now} = \infty$
5. While $n_B < n$ and stopping criteria has not been met, do:
   a. Use the updated set of $B$ to define $P_{limit}(B, n_B)$
   b. $P_{limit}(B, n_B)$ by Chambolle-Pock algorithm[56]
   c. For each structure $s_t \in S$, calculate the Lagrange multiplier as $z_{s_t}$, whose size is the number of voxels in $s_t$
   d. Define vector $v_b$ for all beams ($b \in B_{all}$)
   e. For each beam $b_j \in B_{all} \backslash B$ :
      i. For each structure $s_t \in S$:
         1. Calculate $D_{b_j s_t}$, the dose matrix of beam $b_j$ for $s_t$
         2. $v_{b_j} += D_{b_j s_t} z_{s_t}$
      ii. Set $r_{b_j} = -v_{b_j}$
   f. If $r_b \leq 0 \qquad \forall\, b \in B_{all}$
      i. Stop the algorithm. The solution is optimal.
   g. Otherwise:
      i. Normalize $r$ values by Calculate fitness vector $f$ for all beam with equation (9)
      ii. $\tilde{b} = argmax(f)$,
      iii. Update the associated element in array A: ($\alpha_{\tilde{b}}$) to 1, $A_{\alpha_{\tilde{b}}} = 1$
      iv. $n_B = n_B + 1$
6. Return A

---

## II.C.2   FMO Objective Function of BOO Problem

In step 5.b of **Algorithm 2.** , the FMO should be solved for currently selected beam orientations. As a feasibility study, we define $F(y)$ to be a quadratic penalty function, although other objective functions may be used for the FMO problem as well:

$$F(y) = F\left(\begin{bmatrix} y_{s=s_1} \\ \vdots \\ y_{s=s_T} \end{bmatrix}\right) = \sum_{s \in S} \frac{w_s^2}{2} \|y_s - p_s\|_2^2 \tag{16}$$

where $s$ is the structure index and $S$ is the set of all structures. Each structure has a user-defined structure weight, $w_s$, and a prescription dose, $p_s$. To quickly solve this optimization objective function, we employed the Chambolle-Pock algorithm on a GPU. Technically, many optimization algorithms may be used to solve the FMO problem, but Chambolle-Pock can solve large scale problems quickly and does not need to solve a system of linear equation at each iteration, which makes it a suitable candidate for GPU implementation.

## II.D   Data Generation, Training, and Evaluation

We used images from 70 patients with prostate cancer, each with 6 contours: PTV, body, bladder, rectum, left femoral head, and right femoral head. Additionally, skin and ring tuning structures were added during the fluence map optimization process to control high dose spillage and conformity in the body and to train and test the network. The patients were divided randomly into two exclusive sets: 1. a model development set, which includes training and validation data, consisting of 57 patients; and 2. a test data set consisting of 13 patients.

In this work, we generated scenarios (inputs and outputs) in two ways to create a singular comprehensive training set. The first method runs the CG algorithm in full to select all 5 beams sequentially, using **Algorithm 2.** . This generates 5 cases of training data for the neural network. The second method starts with a random selection of 1 to 4 beams as part of the selected beam set, and the CG algorithm then is run to select the next beam. This second method serves to allow for network to learn how the CG algorithm compensates for when a suboptimal beam set is introduced as input. This will let the network self-correct its own potential errors as well as allow for a human operator to manually edit any of the earlier beams. For either of these methods, the structure weight selection scheme is the same, which is outlined by the following process:

1. In 50% of the times, a uniform distribution in the range of 0 to 0.1 is used to generate a weight for each OAR separately.
2. In 10% of the times, the smaller range of 0 to 0.05 is used to select weights for OARs separately, with uniform distribution.
3. And finally in 40% specific ranges were used for each OAR: Bladder: [0,0.2], Rectum: [0,0.2], Right Femoral Head: [0,0.1], Left Femoral Head: [0,0.1], Shell: [0,0.1] and Skin: [0,0.3]

The weights range from 0 to 1. This weighting scheme were found to give a clinically reasonable dose, however, the dose itself may not be approved by the physician for that patient.

By using ten instances of the first scenario generation method for each patient, $3500 (= 70 \times 10 \times 5)$ different scenarios are created. Then 60 different scenarios are created for each patient in training and validation data set using the second scenario generation method (12 per each size of the input beam set) leads to $3420 (= 57 \times 12 \times 5)$ additional scenarios, which makes $6920$ different scenarios in total. $6270 (= 57 \times (10 + 12) \times 5)$ unique scenarios of this set are used for training and validation, and 650 scenarios for testing. By using $90°$ rotation, we augmented the total number of scenarios in training and validation data set by 4, so at the end we have $25080$ input scenarios for training and validation.

To train the model, we used a 6-fold validation technique, where all folds had the scenarios of 47 patients as their training set. Folds one through five had 10 validation patients, while the sixth fold had seven validation patients. The validation set for each fold is a rolling set of patients of size 10, which varies from each fold to another.

Folds were trained over 400 epochs of step size 2500. In each fold, the model with the least validation loss was chosen and evaluated over the test set. To evaluate the performance of the deep learning model, each model was tested for one-on-one prediction, the same method used for training. All six models were evaluated on test patient scenarios in 2500 steps, and the average loss function was used to measure their performances.

In the next step, the trained DNN model predicted five beam orientations for all the test scenarios, and their associated FMO problems were solved. Finally, considering only test scenarios, FMO solutions of beam sets generated by CG and predicted by DNN were compared with the following metrics:

- **PTV $D_{98}$, PTV $D_{99}$:** The dose that 98% and 99%, respectively, of the PTV received
- **PTV $D_{max}$:** Maximum dose received by PTV
- **PTV Homogeneity:** $\left(\frac{PTV\ D_2 - PTV\ D_{98}}{PTV\ D_{50}}\right)$ where PTV $D_2$ and $D_{50}$ are the dose received by 2% and 50%, respectively, of PTV
- **Van't Riet Conformation Number:** $\left(\frac{(V_{PTV} \cap V_{100\%Iso})^2}{V_{PTV} \times V_{100\%Iso}}\right)$ where $V_{100\%Iso}$ is the volume of the isodose region that received 100% of the dose

- **R₅₀:** $\left(\frac{V_{50\%Iso}}{V_{PTV}}\right)$ where $V_{50\%Iso}$ is the volume of the isodose region that received 50% of the dose

## III. **Results**

To emphasize the effectiveness of the proposed method, we only considered five beam treatment plans for patients with prostate cancer in this study. It has been shown that the selection of beam orientations is more important for plans with smaller numbers of beams.[19]

To train and evaluate the DNN model, we used a 6-fold validation technique, as described in section II.D, where models in all six folds were trained over 400 epochs, each with 2500 steps. It took approximately three days for each fold to complete its training. To compare the performance of the proposed network architecture with a more conventional architecture, we replace the five stages of the third block with five fully connected layers, this updated network is called as DNN_Full throughout the paper. DNN_Full is trained with the same data and the same format as the proposed DNN; using 6-fold cross validation technique, 400 epochs and 2500 steps. **Figure 3** represents the progress of the average and standard deviation of the training and validation loss functions (MSE) during the training process of all folds for proposed DNN and DNN_Full networks.

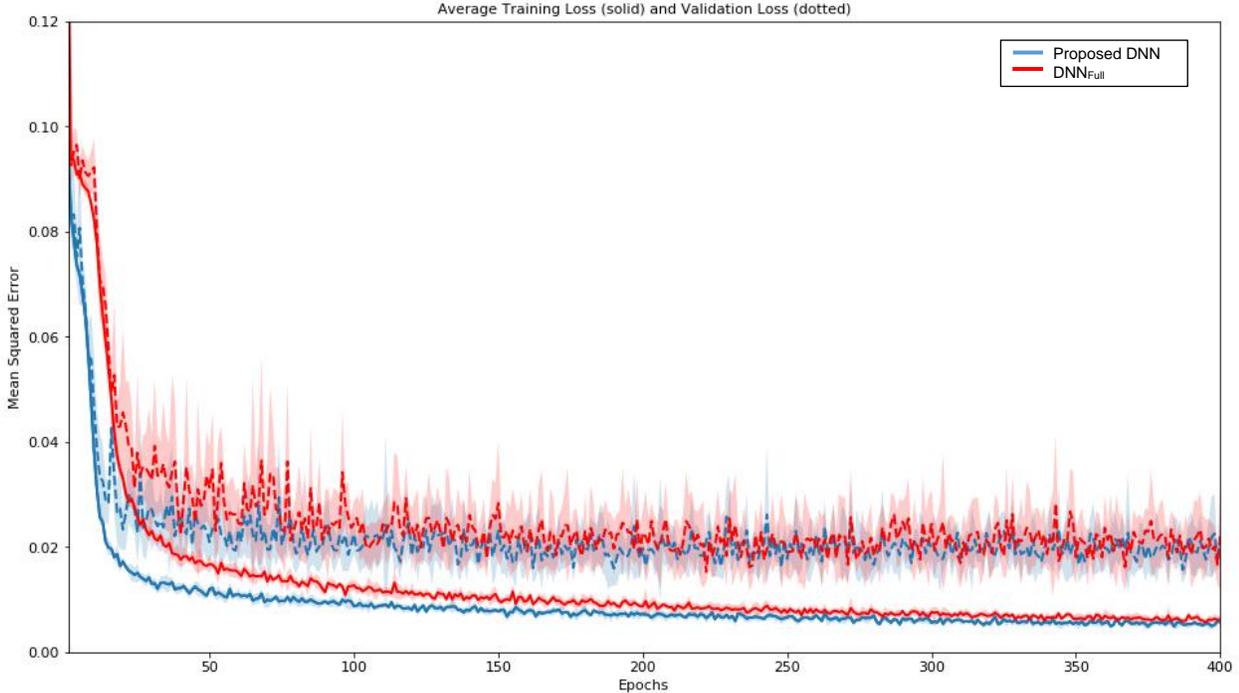

**Figure 3.** *Average Training (Solid) and Validation (Dotted) Loss function (MSE) values for 6-fold Cross-validation for Proposed DNN(Blue) and DNN_Full (Red) networks*

Trained models at the end of each epoch were saved. Therefore, by the end of the training, each fold had generated 400 trained models to choose from, among which the trained model with the least validation loss function was labeled as *Best* and chosen for future analysis. To show the differences between the *Best* and the last trained model (labeled *Last*) within each fold, the values of the training, validation, and testing loss functions (MSEs)—each evaluated on associated training and validation data sets and the general test data set—are presented in **Table 3**. The epoch number associated to the validation best of all folds are also provided in this table. Each fold is trained for 400 epochs, so the last loss function values in this table refers to loss function value of the epoch number 400.

**Table 3.** *Average Training, Validation, and Test loss functions (MSE%) of the proposed DNN network for selected epochs*

| Fold # | 1 | 2 | 3 | 4 | 5 | 6 | Total [d] |
|---|---|---|---|---|---|---|---|
| Train Best [e] | 0.50% | 0.63% | 0.62% | 0.68% | 0.53% | 0.78% | 0.62 ± 0.09% |
| Validation Best [f] | **0.91%** | 1.04% | 1.06% | 1.06% | 1.12% | 1.06% | 1.04 ± 0.06% |
| Test Best [g] | **1.39%** | 1.44% | 1.50% | 1.30% | 1.39% | 1.64% | 1.44 ± 0.11% |
| Training Last [e] | 0.62% | 0.52% | 0.63% | 0.63% | 0.53% | 0.50% | 0.57 ± 0.06% |
| Validation Last [f] | 1.92% | 1.17% | 2.54% | 2.49% | 2.62% | 2.18% | 2.15 ± 0.50% |
| Test Last [g] | 1.59% | 1.54% | 1.73% | 1.79% | 1.68% | 1.65% | 1.66 ± 0.08% |
| Best Epoch Number | 374 | 303 | 189 | 272 | 387 | 165 | |

**Table 4**. shows the train, validation and test loss function value of the $DNN_{Full}$ network. To compare its performance with proposed DNN the total values of both models, proposed DNN and $DNN_{Full}$ are presented side by side in this table. Except for the validation loss function of the best epochs, the loss function value of the proposed method is better (smaller) than $DNN_{Full}$. The chosen trained model of the proposed DNN is from fold 1 and epoch 374, with MSE = **0.91%,** while the best performance for $DNN_{Full}$ is from fold 1 and epoch 292, with MSE = **0.67%**. Although $DNN_{Full}$ has lower **Validation Best** MSE values compares to proposed DNN, it is outperformed by proposed DNN when comparing the **Test Best** MSE values in every fold. In fact, the superiority of the proposed DNN is so much that its largest **Test Best** MSE value (**1.64%**) is smaller than the lowest value of **Test Best** MSE loss of $DNN_{Full}$, even for the best-trained model of $DNN_{Full}$, which has a lower validation loss function.

**Table 4.** *Average Training, Validation, and Test loss functions (MSE%) for Best and Last epochs of the trained $DNN_{Full}$*

| Fold # | 1 | 2 | 3 | 4 | 5 | 6 | Total $DNN_{Full}$ | Total Proposed DNN |
|---|---|---|---|---|---|---|---|---|
| Train Best | 0.768% | 0.857% | 0.630% | 0.726% | 0.596% | 0.491% | 0.678±0.120% | **0.62 ± 0.09%** |
| Validation Best | **0.671%** | 0.990% | 0.806% | 1.082% | 1.177% | 1.110% | **0.973±0.179%** | 1.04 ± 0.06% |
| Test Best | **1.780%** | 2.482% | 2.524% | 2.751% | 1.956% | 3.520% | 2.502±0.565% | **1.44 ± 0.11%** |
| Train Last | 0.572% | 0.525% | 0.564% | 0.609% | 0.619% | 0.84% | 0.621±0.101% | **0.57 ± 0.06%** |
| Validation Last | 1.910% | 1.810% | 1.049% | 2.270% | 1.896% | 4.47% | 2.234±1.065% | **2.15 ± 0.50%** |
| Test Last | 1.427% | 2.977% | 2.196% | 2.540% | 2.146% | 2.27% | 2.259±0.466% | **1.66 ± 0.08%** |
| Best Epoch # | 262 | 238 | 367 | 377 | 276 | 349 | | |

**Figure 4** shows three of the worst $DNN_{Full}$ and proposed $DNN_{Full}$ predictions. Scenario presented in right side of this figure, shows one of the worst predictions for our proposed method (MSE = 9.39%). In this example even though the mean squared value of our proposed method is high, but it can be seen that it captures the trend of the fitness values and therefore chose the beam angle which is very close to CG selected beam. Example to the left is one scenario where $DNN_{Full}$ performed poorly in prediction, not only $DNN_{Full}$ did not choose any local optimum point in the prediction, it faild to recognize the total pattern of the fitness values, and chose a beam angle with minimum true fitness value. The performance of our proposed model with different number of input beam is provided in

**Table 5**. The average MSE value of the cases with four input beams are higher than other cases. We believe more scenarios with four input beams can improve the output of the model in this case. The study of the performance of the network with more input data (more scenarios) is considered for our future research.

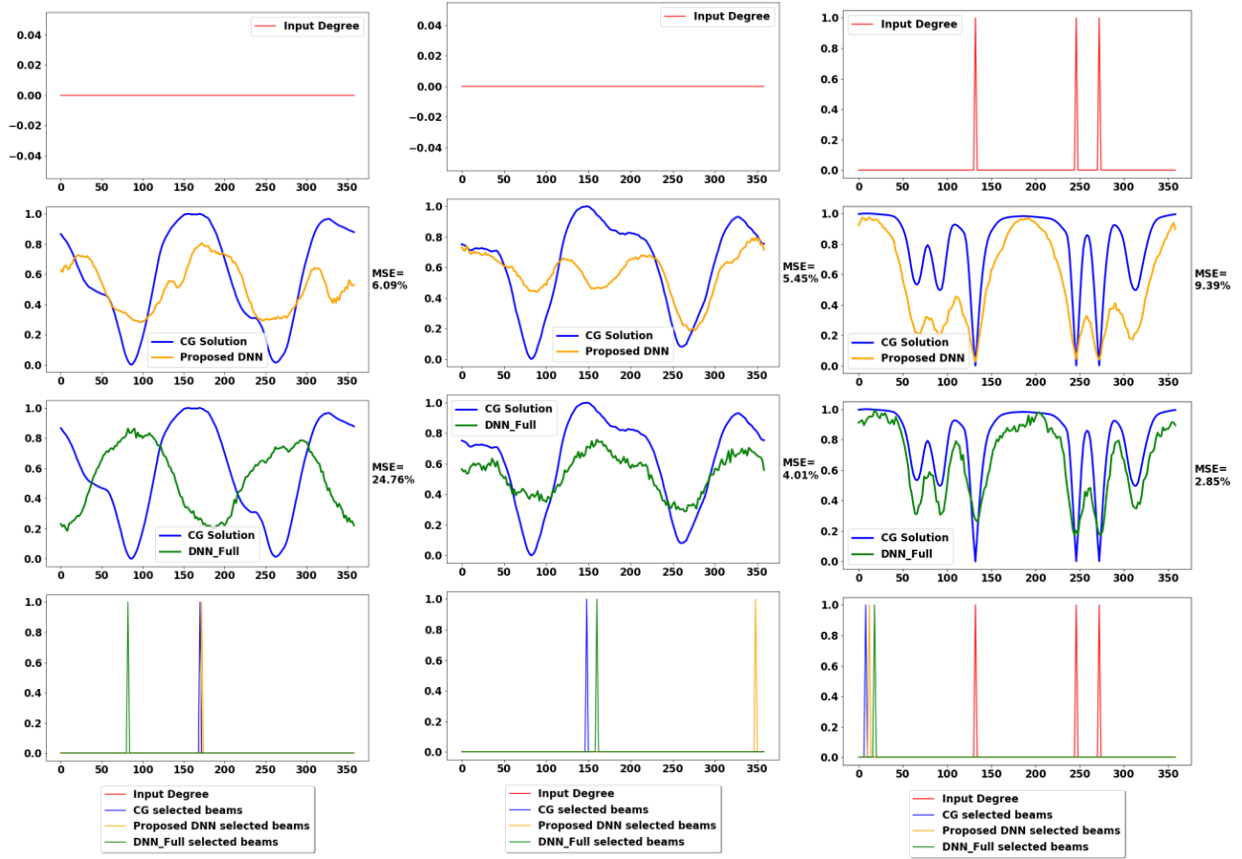

**Figure 4.** *Three Example of worst fitness value prediction by Proposed DNN and DNN_Full.*

**Table 5.** *MSE values of test scenarios with different sizes of input beam set*

|        | The Size of the Set of Input Beams | | | | | |
|--------|--------|--------|--------|--------|--------|--------|
|        | *0* | *1* | *2* | *3* | *4* | *Average* |
| *Fold 1* | 1.67% | **0.58%** | 1.16% | 1.39% | 1.91% | 1.34±0.46% |
| *Fold 2* | 2.17% | **0.35%** | 1.17% | 1.66% | 2.01% | 1.47±0.66% |
| *Fold 3* | 2.12% | **0.44%** | 1.22% | 1.80% | 2.50% | 1.62±0.72% |
| *Fold 4* | 1.56% | **0.33%** | 1.38% | 1.54% | 1.98% | 1.36±0.55% |
| *Fold 5* | 1.76% | **0.32%** | 1.42% | 1.83% | 2.63% | 1.59±0.75% |
| *Fold 6* | 1.69% | **0.53%** | 1.31% | 1.66% | 2.17% | 1.47±0.54% |
| *Total* | **1.83±0.23%** | **0.43±0.10%** | **1.28±0.10%** | **1.65±0.15%** | **2.20±0.27%** | **1.48±0.63%** |

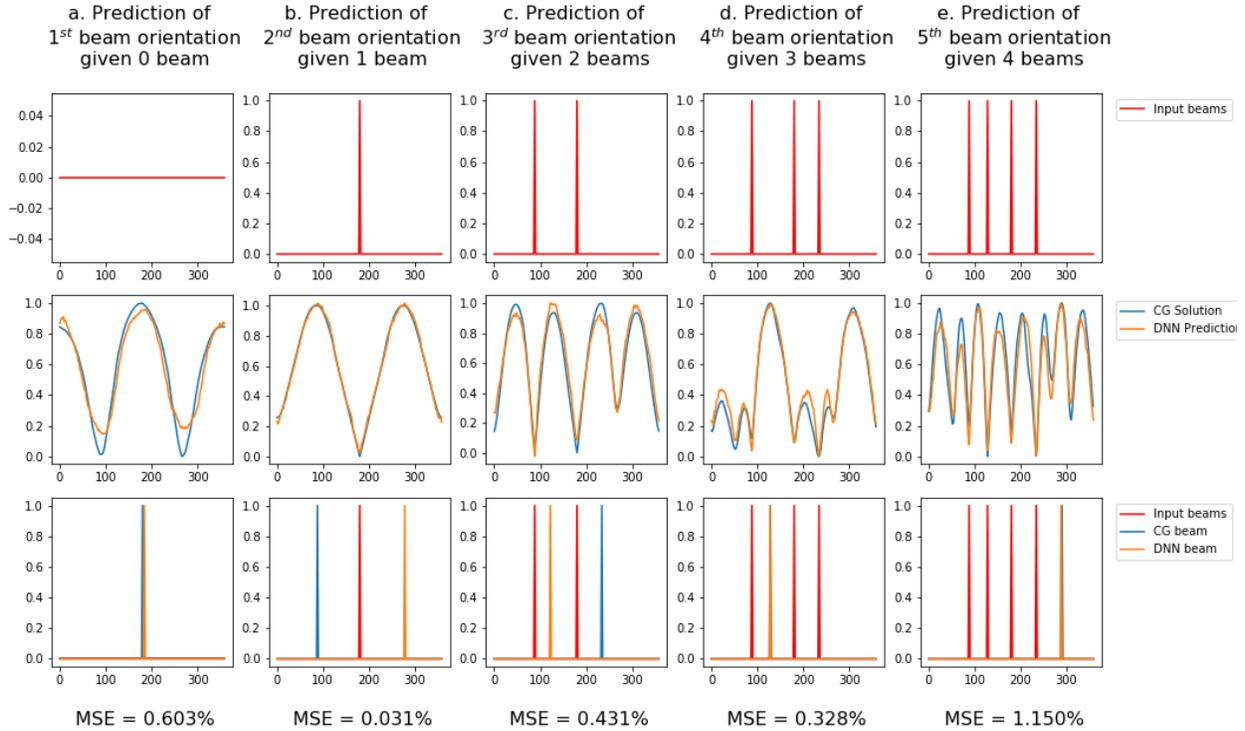

**Figure 5.** *DNN prediction vs CG results, and selected beam orientations.*

The DNN predicted the dual value of each beam that was learned from column generation results; examples of the predicted (DNN) and actual (CG) of the additive inverse of respective dual values are depicted in **Figure 5**. In this figure, the first row indicates the set of already selected beam orientations, or the second input to the network. Graphs in the second row represent the CG and DNN fitness values for a test patient, given the input; and finally, the third row depicts the updated set of beam orientations based on the input beams and argmax of fitness values for CG and DNN separately. After dual values were calculated, the beam orientation with the minimum dual value (maximum in the case of negation) was chosen to be added to the current set of selected beams (B). Since the maximum number of beam orientations needed for this project is five, examples up to the prediction of the fifth orientation are provided in **Figure 5**. Although the DNN predicted values were very close to CG, with 1.44% average MSE value, the DNN may choose a different beam orientation than CG in the presence of multiple local optimums (e.g. **Figure 5.b and Figure 5.c**) or when dual values are very close to each other (e.g. **Figure 5.a** and **Figure 5.e**).

The trained DNN model was called consecutively to predict five beams for each patient. The DNN started with the patient anatomy in the form of target and avoidance matrices (*Input 1*), as mentioned in section 2.1, and a zero array of size 180 (the number of candidate beam orientations), in which each element is associated with one candidate beam (*Input 2*). The DNN took *Input 1* and *Input 2* as its initial inputs and predicted the output (U) array. Then, the index of the beam with the minimum (maximum) U value (additive inverse of U) was determined, and the value of its respective element in *Input 2* changed to one. Then, the DNN took the *Input 1* and the updated *Input 2* to predict another U array, and the index of its minimum was used as the next element updated. This process continued until there were five values of 1 in the *Input 2* array. The FMO problem of the five beam orientations predicted by DNN was solved, and its result was used for further evaluation and comparison with the FMO solution of the five beams selected by CG.

The proposed DNN is trained to have self-correction ability, as mentioned in section II.D. The self-correction helps to rectify the poor prediction of DNN in earlier iteration. **Figure 6** demonstrates an example

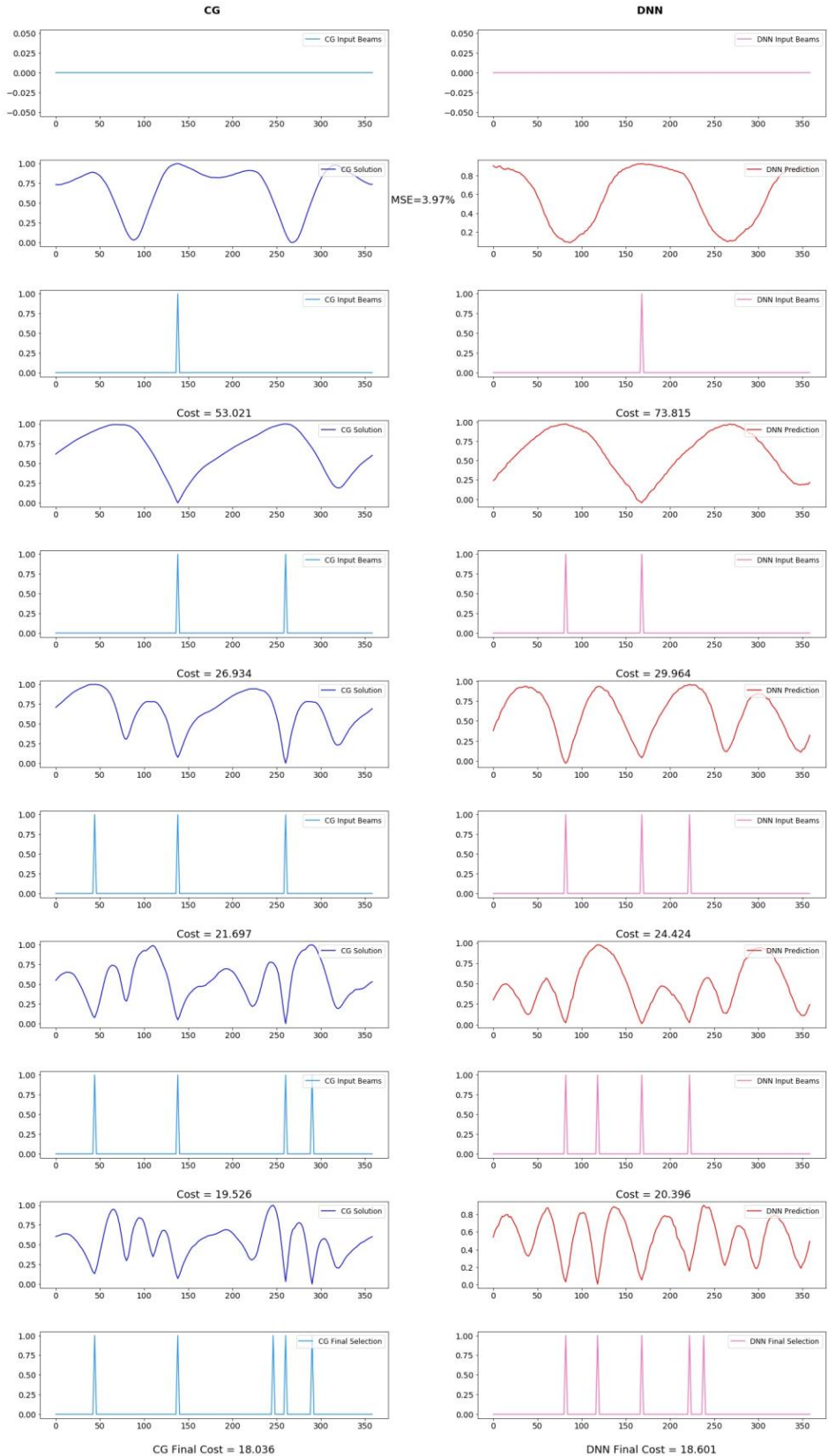

**Figure 6.** *Self-correction of the proposed DNN*

of the self-correction ability of DNN where it makes a poor prediction in the first iteration.

As mentioned earlier, DNN and CG may predict different beam orientations, because there might be several beam orientations within $\varepsilon$ distance of the minimum dual values. Therefore, DNN may select a different beam set than CG, but we expect the quality of the solutions to be almost the same. Both CG and DNN were evaluated on the test data set, and two beam sets related to each algorithm (CG and DNN) were generated and evaluated by the six metrics listed in section II.D. The solutions of the iterative prediction of deep learning model and column generation algorithms and their differences are presented in **Table 6**. Note that the final solutions to DNN and CG are the final dose calculations resulting from solving the FMO problems.

**Table 6.** *Iterative Column Generation and predicted values of test dataset for PTV statistics, van't Riet Conformation Number (VR), and High Dose Spillage ($R_{50}$) in the form of mean value ± standard deviation for six folds and in total*

| | | PTV $D_{98}$ | PTV $D_{99}$ | PTV $D_{max}$ | PTV Homogeneity | VR | $R_{50}$ |
|---|---|---|---|---|---|---|---|
| | CG | 0.977±0.011 | 0.961±0.020 | 0.870±0.059 | 0.069±0.038 | 0.881±0.083 | 4.676±0.888 |
| **Cross-Val 1** | DNN | 0.977±0.011 | 0.960±0.019 | 0.871±0.055 | 0.070±0.039 | 0.863±0.108 | 4.400±0.622 |
| | CG-DNN | 0.000±0.003 | 0.000±0.005 | -0.001±0.029 | -0.001±0.004 | 0.018±0.035 | 0.276±0.378 |
| **Cross-Val 2** | DNN | 0.976±0.012 | 0.960±0.020 | 0.870±0.057 | 0.070±0.038 | 0.879±0.093 | 4.555±0.694 |
| | CG-DNN | 0.000±0.003 | 0.001±0.005 | 0.000±0.031 | -0.001±0.004 | 0.002±0.021 | 0.121±0.372 |
| **Cross-Val 3** | DNN | 0.977±0.011 | 0.960±0.020 | 0.874±0.055 | 0.070±0.038 | 0.873±0.067 | 4.571±0.760 |
| | CG-DNN | 0.000±0.003 | 0.000±0.005 | -0.004±0.028 | -0.000±0.004 | 0.007±0.024 | 0.105±0.441 |
| **Cross-Val 4** | DNN | 0.976±0.011 | 0.960±0.019 | 0.873±0.055 | 0.070±0.038 | 0.867±0.062 | 4.528±0.768 |
| | CG-DNN | 0.000±0.003 | 0.001±0.005 | -0.003±0.031 | -0.001±0.003 | 0.013±0.055 | 0.148±0.277 |
| **Cross-Val 5** | DNN | 0.977±0.011 | 0.960±0.020 | 0.873±0.056 | 0.071±0.038 | 0.881±0.090 | 4.371±0.692 |
| | CG-DNN | 0.000±0.003 | 0.000±0.005 | -0.002±0.028 | -0.001±0.005 | -0.001±0.015 | 0.305±0.423 |
| **Cross-Val 6** | DNN | 0.976±0.012 | 0.960±0.020 | 0.868±0.061 | 0.070±0.038 | 0.878±0.083 | 4.478±0.612 |
| | CG-DNN | 0.001±0.003 | 0.001±0.005 | 0.002±0.030 | -0.001±0.004 | 0.003±0.030 | 0.198±0.402 |
| **Total** | DNN | 0.977±0.012 | 0.960±0.020 | 0.872±0.057 | 0.070±0.038 | 0.874±0.085 | 4.484±0.698 |
| | CG-DNN | 0.000±0.003 | 0.001±0.005 | -0.001±0.029 | -0.001±0.004 | 0.007±0.034 | 0.192±0.393 |

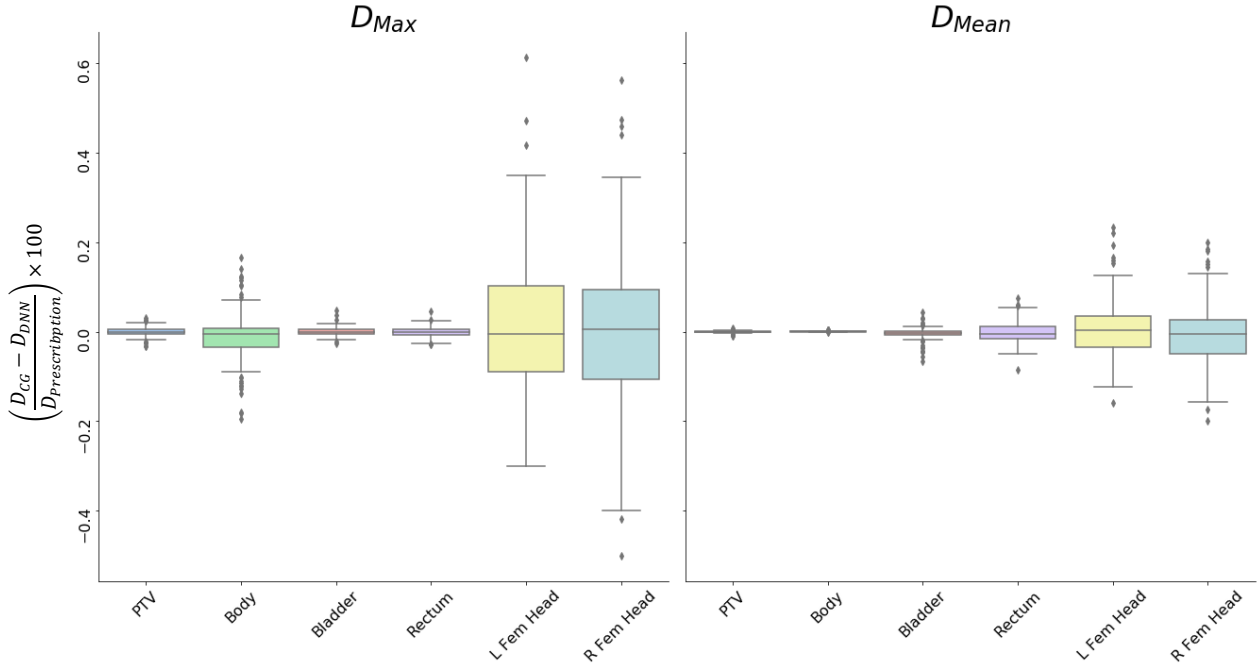

**Figure 7.** *Max and Mean of the dose difference percentage between the final solutions of CG and DNN.*

The maximum and the average dose difference percentages of CG's and DNN's final solutions are presented in **Figure 7**. Prescription dose in the dose difference formulation is considered as one for all OARs and the PTV. In particular, on average the dose differences received by organs at risk were between 1 and 6 percent: Bladder had the smallest average difference in dose received (0.96±1.18%), then Rectum (2.44±2.11%), Left Femoral Head (6.03±5.86%), and Right Femoral Head (5.89±5.52%). The dose received by Body had an average difference of 0.10± 0.10% between the generated treatment plans. Although we compare the results of these two approaches together, these approaches are heuristic and may not result in the optimal solution, so they cannot be used as the true optimal value. Therefore, we cannot use any of these solutions as the ground truth of the problem, and the results presented here are only for the sake of comparison.

Examples of the final FMO solutions derived from beam orientations selected by CG and DNN are presented in **Figure 8**. Structure weights affected the beam selection and final FMO solutions. To understand their impacts, the weights used in the model are presented in the first line of **Figure 8**. The Dose-Volume Histogram (DVH) graph of the treatment plans resulting from solving the FMO for each set of solutions is presented in **Figure 8.a**, and dosewashes of CG and DNN treatment plans are provided in **Figure 8.b** and **Figure 8.c**, respectively. In this example, three of the selected beams are within 8° distance of each other; in these cases, the prediction error happened in the shape of the tip of the graphs, as shown in **Figure 5.e**. The dose influence matrices of beams in close vicinity are expected to be very similar, or at least not dramatically different from each other. In such cases, the neighborhood of the optimal point was correctly predicted, and the error happened in the distribution of elements in that neighborhood. The other two beams in CG and DNN were almost on the opposite side of each other (180° differences), 212° in the CG solution vs 36° in the DNN solution, and 354° in the CG solution vs 174° in the DNN solution. We believe that, because of the symmetric anatomical features of the patient's body, selection of two opposite beams may have the same effect on the body and may result in multiple local optimums very close to the global optimum; an example of this case can be seen in **Figure 8.c.** Hence, beam orientation selections of CG and DNN were different in

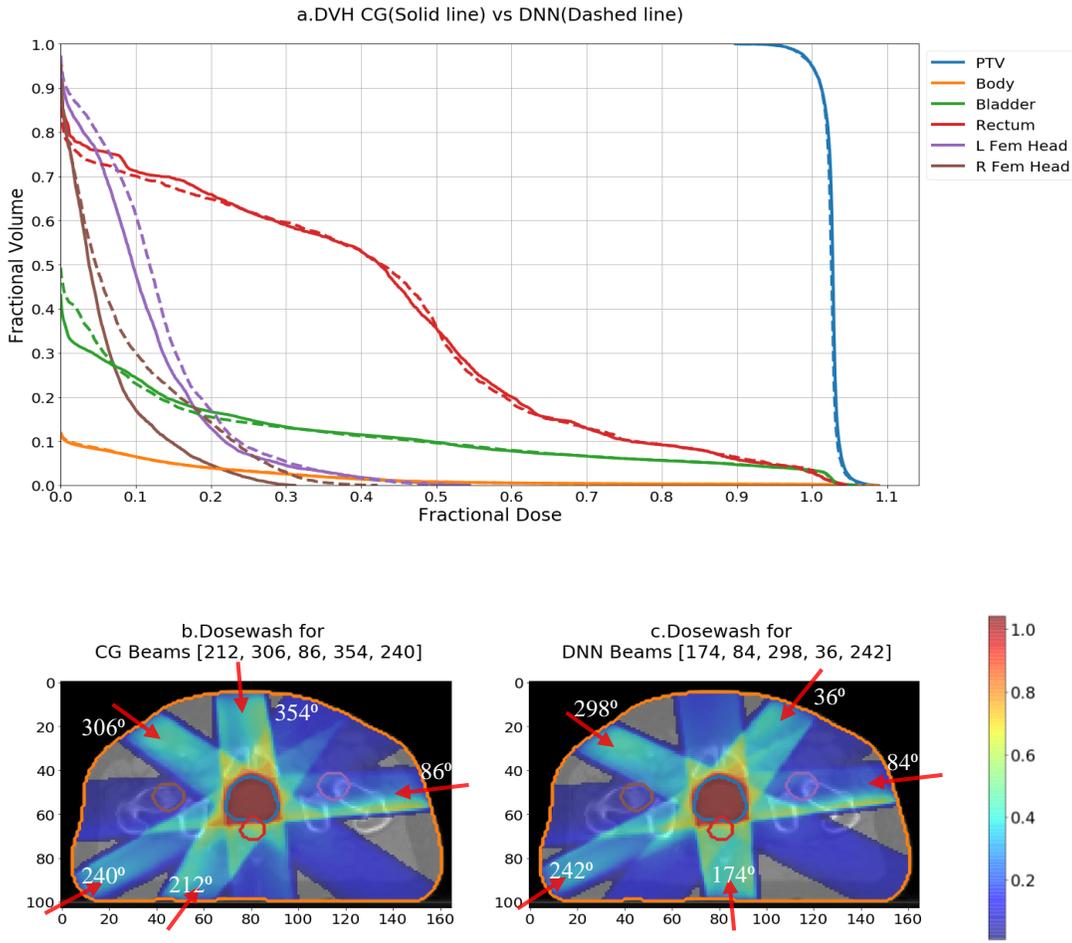

Structure Weights:  PTV/1.00  Bladder/0.06  Rectum/0.039  L Fem Head/0.07  R Fem Head/0.001  Shell/0.029  Skin/0.097
CG_Cost:  10.30  vs  DNN_Cost:  10.38

|  | PTV $D_{98}$ | PTV $D_{99}$ | PTV $D_{max}$ | PTV Homogeneity | van't Riet Conformation Number | $R_{50}$ |
|---|---|---|---|---|---|---|
| **CG Values** | 0.984 | 0.970 | 0.898 | 0.065 | 0.915 | 3.598 ± 0.000 |
| **Pred Values** | 0.982 | 0.970 | 0.893 | 0.068 | 0.908 | 3.664 ± 0.000 |
| **CG-Pred** | 0.002 | -0.000 | 0.005 | -0.003 | 0.007 | -0.066 ± 0.000 |

**Figure 8.** *CG and DNN solutions comparison. Information regarding the structure weights and final objective values of CG and DNN solutions are listed at the top of the figure. The figure in the first row (a.) shows DVH of CG solution (Solid) vs DNN solution (Dashed), and figures in the second row (b. and c.) illustrate the dosewash of the selected beams by CG (left) and DNN (right).*

**Figure 8**, but the DVH curves are very similar, which means that plans generated by CG and DNN solutions affect the patient's body the same, which is expected from the DNN solution.

**Figure 9** illustrates the percentages of scenarios in which CG found a better solution compared to DNN and vice versa, for various $\bar{\varepsilon}$, where $\bar{\varepsilon}$ is an acceptable error margin for equivalent solutions. Two solutions are considered equivalent in this study if and only if the values of their cost functions are within $\bar{\varepsilon}$ distance of each other. CG Better means $\varepsilon_{CG} < \bar{\varepsilon}$, and DNN better means $\varepsilon_{DNN} < \bar{\varepsilon}$; note that $\varepsilon_{DNN} = 0$ and $\varepsilon_{CG} = 0$ are excluded from the comparisons. The calculations of $\varepsilon_{DNN}$ and $\varepsilon_{CG}$ are provided by the following equations:

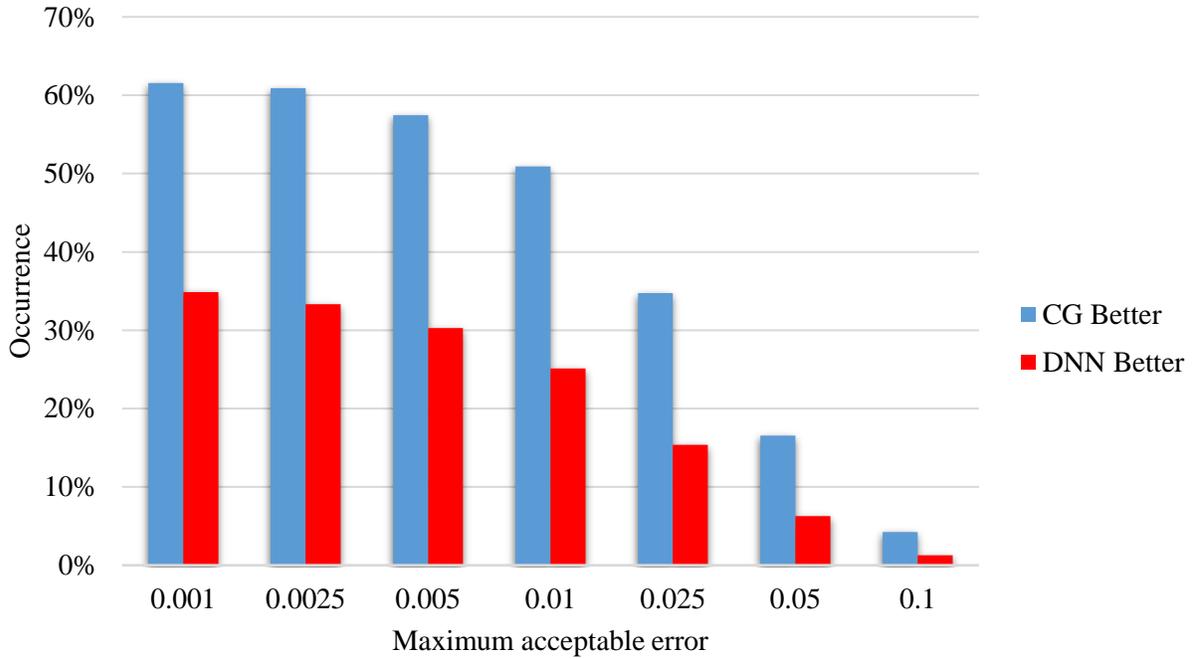

**Figure 9.** *Comparison of FMO Cost Functions for beams selected by CG vs DNN.*

$$\varepsilon = \frac{CG_{Cost} - DNN_{cost}}{CG_{Cost}}$$

$$\varepsilon_{CG} = \varepsilon^-$$

$$\varepsilon_{DNN} = \varepsilon^+$$

(17)

As shown in **Figure 9**, the higher the value of $\bar{\varepsilon}$, the less the superiority of CG and DNN plans. Although on average CG found better solutions, in 35% to 1% of the test scenarios with error thresholds ($\varepsilon$) in the range of 0.001 to 0.1, respectively, DNN got better results.

## IV.  Discussion

The optimal BOO solution can be defined as a set of beam orientations that produces the *optimal treatment plan*. However, there is no globally accepted definition of the optimal treatment plan in mathematical terms, despite its simple literal definition.[h] Hence, it is important to consider the flexibility of the treatment planning objective function in defining the BOO problem. A comprehensive mathematical description of the BOO problem, independent of the treatment planning objective function, is presented by Ehrgott et al.[76] This study presents a supervised deep learning neural network model (DNN) that learns from a CG algorithm. CG operates by first generating fitness values for the first beam, and then running an FMO with the selected beam. CG then uses the "optimized" dose from the FMO to calculate fitness values to select the second beam. This alternation between the FMO and the fitness value calculation is the key for the effective operation of CG, but is also the reason that CG is slow. The proposed DNN network, while not necessarily solving an FMO at each iteration, does learn the fitness values that have been calculated using  the FMO information.

The DNN inputs are the anatomical information, PTV and OARs, and a set of already selected beam orientations. The output is an array of predicted dual values (the changes in the objective function resulting from adding a beam orientation, in other words, probability of success). The proposed DNN learns from a column generation (CG)-based algorithm to predict dual values. Then, the beam orientation with the most negative dual value is selected to join the set of already selected beams ($B$). The proposed CG algorithm is a greedy algorithm that selects a set of beam orientations among a set of candidate beams by iteratively calling the Chambolle-Pock algorithm, calculating dual values, then finding the beam orientation with the best possible outcome for the current state of the problem (the process in each iteration is referred to as CP-DL-ST). The DNN learns its behavior from the full implementation of one iteration of CG, the CP-DL-ST processes; therefore, the DNN also tries to predict the best beam orientation that can be added to the current state of the problem. Note that as a greedy algorithm, CG does not guarantee global optimality, but it has been shown to produce superior results for treatment planning problems in existing literature[22,49-55].

The main justification to use convolution layers is that convolutions make a locality assumption that the adjacent data is related to each other. We make the assumption/constraint that the fitness value at one degree is similar to the fitness value of its neighbor beam (e.g. a beam 0 degrees and a beam 2 degrees are more similar than 0 degrees and 8 degrees). A fully-connected architecture can theoretically learn how to do this as well (CNNs are a subset of Full connections), but typically requires more data to learn that relationship, and is an issue for our limited data issue. This is also very apparent in **Figure 4**, where the fully connected architecture produces more jagged prediction of the fitness values, leading to more volatile beam angle selections.

The large number of candidate beams for IMRT treatment planning (e.g., 180 candidate beams for a coplanar geometry with 2-degree separation and 1162 candidate beams for a noncoplanar geometry with 6-degree separation) creates a very large solution space that is difficult to fully explore in a feasible amount of time. In such cases, CG performs efficiently by limiting the solution space and exploring subareas with the highest potential first. CG limits the solution space by dividing the original problem into sub-problems, in our case $P_{limit}$, and solving selected sub-problems. The Chambolle-Pock algorithm is used to solve $P_{limit}$, and as iterative optimization approach, it will converge to $P_{limit}$'s optimal solution, since $P_{limit}$ is convex.

Using the DNN only once is good for predicting one (next) beam orientation, but in BOO problems, the selection of N beams is of interest. Therefore, to use the DNN to find N beams, DNN should be called N times or in N iterations. The DNN tries to optimize and then update the current state of the problem, and the updated state is used in the next iteration. By the end, N different states of the problem will have been created and optimized to find N beam orientations. An algorithm with an iterative structure that optimizes a limited problem in each iteration is considered a greedy algorithm; greedy algorithms do not guarantee the optimality of the final solution, and they usually end up with a local optimum solution. Since DNN is a greedy algorithm, its solution may not be an optimal solution and potentially could be improved.

As a feasibility study, we trained the model to predict up to 5 beams. However, the model architecture is designed to be fully extendable to any number of beams, and having accurate predictions would require training the model in these higher beam spaces. Technically, the current model can predict a 6th, 7th, etc. beam, but since it has not trained on this area, the prediction error is expected to increase. Since column generation is a greedy algorithm, its selection of the 5th beam is always the same regardless if algorithm stops at 5 or at 50 beams. Therefore, our 5 beam plan would not be affected by training the neural network to predict more beams.

The proposed DNN is a very fast algorithm that only needs patient images as an input, and it can provide a solution with a good quality in a matter of seconds, while conventional algorithms take hours. For example, predicting five beams with the DNN takes at most 1.5 seconds, and calculating dose influence matrices of five beams and solving the respective FMO problem will take approximately 26 minutes; five minutes for dose calculation per beam and 20 seconds solving the FMO problem. By contrast, calculating dose influence matrices for all possible beams (in our case 180 beams) takes around 15 hours, then selecting five beams using CG and solving FMO iteratively takes 549 ± 131 seconds (varies for different patients). Note that to solve the proposed CG method, the dose influence matrices of all beamlets of all possible beams should be loaded to the computer, which will slow down the process, compares to the time needed to solve five FMO problem consecutively (around 100 seconds). Therefore, even though the DNN's final solution may not be optimal, because of its speed, independence from dose influence matrix calculations, and relatively good solution quality, it can be used independently to create treatment plans with high quality. Or it can be considered as a possible treatment plan for future analysis for various sets of OAR weights and patient images.

Although it is not fully clear what the acceptable error margins are for clinical acceptance. CG already outperforms manually beam selected plans[50]. Also, for these to be used in the clinic, the dosimetrist will be going through many trial-and-error iterations to obtain an acceptable plan. Keep in mind that this "error" is how for the DNN-beam selection strays from the CG plan, but not error in the calculated dose and delivery sense. If the DNN produces a plan different from CG, but the dosimetrist prefers that plan, then they may go with the DNN plan.

As volumetric modulated arc therapy (VMAT) has become increasingly popular because of its high plan quality and efficient plan delivery,[77,78] clinical applications of new methods for standard IMRT treatment planning processes, including BOO, may seem less appealing. However, with the advent of highly noncoplanar plans, such as $4\pi$ radiotherapy[22,50] and Station parameter optimized radiation therapy (SPORT),[79,80] finding a fast method for beam orientation selection is very useful and is of interest. This study examines a neural network's ability to learn how an optimization process behaves based on patient anatomical structures and tests this ability on coplanar IMRT treatment planning of patients with prostate cancer; but for a clinical evaluation of the proposed method, we plan to extend the model to noncoplanar treatment planning features. Considering CG's success in solving highly noncoplanar problems,[22,49-55] we expect that the DNN, learning from the CG policy, will also be able to scale up and efficiently approximate the noncoplanar beam orientation optimization solution.

One interesting observation was that, although DNN learns from CG and is expected to have solutions inferior or equivalent to CG's, there were some scenarios in which DNN outperformed CG. There are two possible explanations for this: 1. The proposed CG is a greedy algorithm, and the solution is not necessarily optimal, so other beam selections may result in better solutions. 2. CG and DNN might select different sets of beams because of the presence of multiple global optimums, and while choosing any of the global optimums is expected to have the same impact on the cost function, the affected anatomical feature of a patient is different, which will cause a different solution. This means that the absolute values of the solutions found by CG and DNN are very close together, so there is a high chance that DNN will predict a set of beam orientations whose plan quality is as high as that from the beams selected by CG (with some error), but DNN will predict the beam orientations much faster than CG.

In addition, as presented in **Figure 7**. the dose differences of femoral heads are considerably higher than other organs. This is likely due to the fact that the femoral heads are further away from the treatment isocenter. Although small changes of a beam orientation may have small impact on organs closer to isocenter (such as bladder and rectum) and PTV, they may cause considerable effect on the dose distribution of OARs farther from isocenter (such as right and left femoral heads). Because of that, two sets of beams with small differences, may result in small changes in the final value of the cost function, while the impact on femoral heads are more significant.

Furthermore, users of the proposed DNN have the freedom to choose their own optimization algorithm, because the proposed DNN only needs an optimal solution for each state of the problem, and we believe it can learn from any type of optimization method and is not restricted to the CG procedure. In addition, we believe the proposed method can be independent of the objective function chosen for the treatment planning problem, but this needs to be tested for confirmation.

## V. Conclusions

In this study, we developed a fast method to approximate the solution to BOO problems. Our iterative method uses a supervised deep learning network to predict the beam orientations for the treatment plan, based on the policy generated by the column generation algorithm. Even though we used CG to train the model, the DNN structure is independent of the optimization algorithm, and any selection method can be used to train the model. We demonstrated the feasibilty of the methodology, using prostate patient data planned with 5 beam coplanar IMRT, and we plan to extend this work to noncoplanar beam orientation optimization for 4π radiotherapy in a future study. Our DNN saves a considerable amount of time, while maintaining plan quality similar to that of CG; because the trained DNN only needs the anatomical information of patients—contoured images and structure weights—to select beam orientations and it only requires dose influence matrix calculations for the selected beams (5 in this study) for the final evaluation, unlike CG that requires the computation of dose influence matrices of all potential beam orientations (180 in this study) in addition to patients anatomical data. The DNN predicts the set of five beam orientations in 1.5 seconds; after the FMO is solved, the total time needed for the DNN is five minutes and 20 seconds, which is much faster than the CG method, which needs more than three hours to calculate the dose of every candidate beam. The extreme speed with which the neural network approximates the BOO solution makes it a viable model for clinical application.

## Acknowledgment

The authors would like to thank Jonathan Feinberg for editing the manuscript. This work was sponsored by NIH grant No. 1R01CA237269-01 and Cancer Prevention and Research Institute of Texas (CPRIT) (IIRA RP150485, MIRA RP160661).

---

[a] $x$ and $y$ are complementary, if $x \lor y = 1, x \land y = 0$

[b] With the help of Latex package in: https://github.com/HarisIqbal88/PlotNeuralNet

[c] Using all possible beam orientations is expensive and very time-consuming, and it reduces the quality of the delivered plan.

[d] Mean $\pm$ Standard deviation of loss function (MSE%) among all folds

[e] Train Best (Last): The training loss function of the epoch with minimum validation Loss (the last training epoch) in each fold

[f] Validation Best (Last): The validation loss of the epoch with minimum validation loss (the last training epoch) of each fold

[g] Test Best (Last): MSE of the prediction value, using the training model of the epoch with minimum validation loss (the last training model) of test dataset compared to the CG values.

[h] The plan that causes the maximum damage to the tumor(s) and the minimum to zero damage to healthy tissues and organs at risk.